\documentstyle[11pt,newpasp,twoside,epsfig]{article}

\begin{document}

\title{Microlensing extended stellar sources}
\author{H M Bryce and M A Hendry}
\affil{Department of Physics and Astronomy, University of Glasgow,
Glasgow, G12 8QQ, UK}

\begin{abstract}
We have developed a code to compute multi-colour microlensing
lightcurves for extended sources, including the effects of limb
darkening and photospheric star spots as a function of spot
temperature, position, size and lens trajectory. Our model also
includes the effect of structure within the spot and rotation of the
stellar source.

Our results indicate that star spots generally give a clear signature
only for transit events. Moreover, this signature is strongly
suppressed by limb darkening for spots close to the limb -- although 
such spots can be detected by favourable lens trajectories.

\end{abstract}

\section{Introduction}
The amplification for a point source microlensing event is 
dependent only on the projected separation between lens and 
source. However, if the source size is comparable to the Einstein 
radius and the projected separation is small, it is necessary to 
treat the source as finite, with the amplification being calculated 
as an integral over the source star. Thus, the microlensing 
lightcurve also contains information about the source surface 
brightness profile.

Finite source effects were considered by Gould~(1994) for the 
case in which a lens \emph{transits} the star, allowing measurement 
of the lens proper motion and thus distinguishing between MC 
and Galactic MACHOs. Witt and Mao~(1994) showed that the magnification 
profile of an extended source with a constant surface brightness 
profile was significantly different from the point source treatment 
for a 100 $R_{\sun}$ source microlensed by a 0.1 $M_{\sun}$ MACHO. 
The ability to recover the lens mass, transverse velocity and the 
source size are considerable assets in extended source microlensing, 
however rare the events may be.

The opportunity to use such events to provide information about the
source star has also been investigated. Simmons \emph{et al.}~(1995) 
modelled the polarisation signature of an extended source 
microlensing event, assuming a pure electron scattering grey model 
atmosphere. They showed that including polarisation information 
significantly improves the accuracy of estimates of the source radius, 
as well as better constraining the radial surface brightness profile 
of the source. Gould~(1997) suggested the use of extended source 
events to determine the rotation speed of red giants. The use of
extended sources to estimate source limb darkening parameters has 
also been discussed by several authors (cf. Hendry 
\emph{et al.} 1998; Valls-Gabaud 1998). Such an event produces a 
chromatic signature as the lens effectively 
\emph{sees} a star of different radius in different photometric 
colour bands. This provides an unambiguous signature of 
microlensing, as opposed to that from a variable star (Valls-Gabaud 
1998).

\section{Microlensing spots}

Recent literature has turned attention to the microlensing of
extended sources with non-radially symmetric surface brightness 
profiles (cf. Hendry \emph{et al.} 2000; Heyrovsk\'{y} and Sasselov 
2000 (hereafter HS00); Kim \emph{et al.} 2000). Microlensing provides an almost
unique probe of such features: doppler imaging is inapplicable to
late type stars, due principally to their long rotation periods, 
and direct optical imaging of stellar photospheres
remains beyond the scope of current technology for all but a
handful of stars. There is, nonetheless, already direct evidence 
for the presence of a large `hot spot' on $\alpha$ Orionis, which
shows strong limb darkening (Uitenbroek \emph{et al.} 1998). 
Hot spots due to convection cells are likely to be present on the 
photospheres of red supergiants, and may be 
associated with non-radial oscillations. \emph{Gravitational imaging} 
of such features on late type giants and supergiants would provide 
valuable constraints on stellar atmosphere models, placing limits 
on e.g. the form of the limb darkening law and the role of rotation.

While one can, of course, also consider the decremental effect of a
\emph{cool} spot, it seems pragmatic -- in terms of 
spot detectability -- to consider the impact of a hot spot on
a cool, massive star. HS00  have demonstrated that hot central spots with a 
radius of $0.2R_{*}$ 
can produce a change in the microlensed flux $\geq 2\%$. We have 
improved their model in several important respects.
\begin{itemize}
\item{We have incorporated the geometric foreshortening of circular 
spots on the photosphere close to the stellar limb, instead of placing 
circular disks on the star. Circular spots close to the limb appear
increasingly elliptical.}
\item{We have computed spot detectability both in terms of the percentage 
deviation from the unspotted flux and the goodness of fit 
of realistic, sparsely sampled observations -- with magnitudes and errors
appropriate to current MC and Bulge data -- to the model of an 
unspotted star.}
\item{We have considered linear, logarithmic and and square root 
limb darkening laws, with band-dependent LD coefficients from LTE 
stellar atmosphere models (Claret and Gimenez 1990) for U to K bands.}
\item{Finally, we have considered the observational signatures of 
multiple spots, spots with temperature structure and spots on rotating 
stars -- features which make our model somewhat more realistic.}
\end{itemize}

\begin{figure}
\plotfiddle{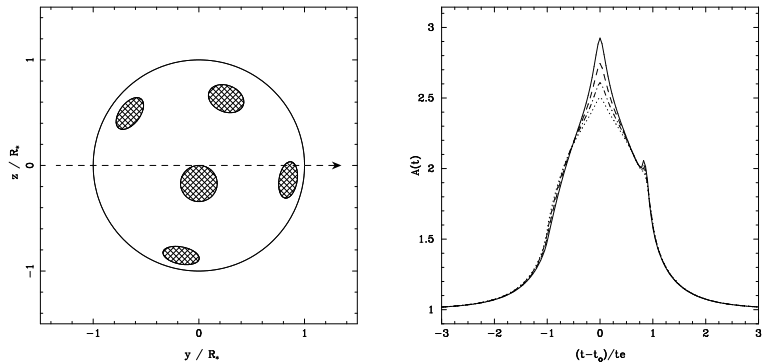}{5cm}{-90}{40}{40}{-150}{200}
%\plotone{helen1.ps}
%\centerline{\epsfig{file=helen1.ps,angle=-90,width=8cm}}
%\centreline{\psfig{file=helen1.ps,angle=-90,width=9cm}}
\caption{\label{one}\small{The microlensing light curve produced by 5
hot (+800K) circular spots on a 4000K $\log g = 2$ star with stellar 
radius equal to the angular Einstein radius (AER) of the lens,
being lensed at zero impact parameter. Note that the spots close to the
limb appear foreshortened and do not contribute significantly
to the microlensed flux. The unbroken, dashed, dot-dashed 
and dotted lines represent B, V, R and I bands respectively.}}
\end{figure}

\section{Results}

We have investigated the microlensing signatures of a range of
spot parameters: radius $5\deg$, $10\deg$, $20\deg$; spot 
temperature relative to stellar effective temperature ranging from
$+200$ K $\rightarrow$ $+1000$ K and $-200$ K $\rightarrow$ $-1000$ K;
spot geometry -- i.e. varying the spot position on photosphere.

We summarise first our qualitative conclusions.
As in HS00 our results show that central hot spots are detectable. 
Furthermore spots close to the limb can also be detected if the lens passes 
close to the feature, especially at minimum impact parameter (see Figs. 3
and 4 below). As the impact parameter increases, the central area where 
spots are detectable contracts, and only very large spots will produce 
detectable signals if the lens does not transit the source. A spot 
close to the limb produces only a small signal, as it is suppressed 
by the geometric foreshortening and the effects of the limb darkening. 
The stellar radius also affects the detectability of spots; however even 
on a small star ($R_{*} \sim 10^{-5}$ AER) a plausible hot spot can be 
resolved at small impact parameter. The effect of a spot is clearly 
\emph{localised} and of short duration, however. For example, a central 
$10\deg$ spot will produce a significant deviation for about $15\%$ of the 
Einstein radius crossing time, with no effect in the lightcurve wings. 
Thus, excellent temporal sampling ($\sim$ hours) is required to detect 
spots -- even for surveys with high photometric precision.

\begin{figure}
\plotfiddle{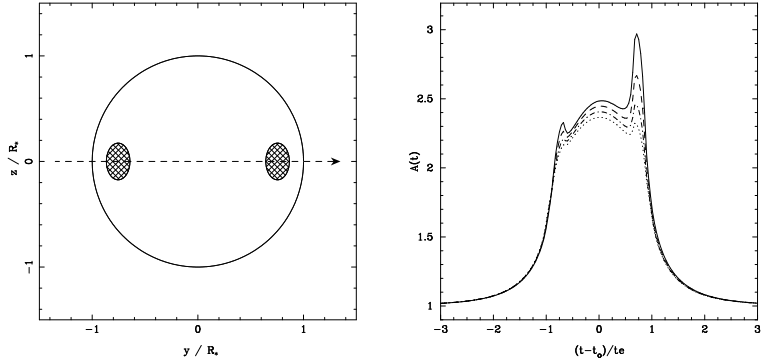}{5cm}{-90}{40}{40}{-150}{200}
\caption{\small{The same star as above, with the same colour bands
represented but with a different maculation. The spot on the left is
as 4500K and produces a 3$\sigma$ signal for 10\% of the lensing event, 
for typical magnitude errors. The spot on the right is at 
5000K and produces a 3$\sigma$ signal for 30\% of the event, as well 
as an higher amplification.}}
\end{figure}

\subsection{Spot temperature and structure}

The percentage deviation from the unspotted flux increases with the
temperature of the spot (see Fig. 2) although the dependence on
temperature is less strong than on spot position. Cool spots produce 
depressions in the lightcurve, but are less 
detectable than a hot spot with the same position, radius and (absolute)
temperature difference. Not only is the light curve deviation smaller
for a cool spot but it is also present over a shorter timescale. Including
the effect of umbrae within spots produces additional small-scale structure 
in the light curve, but differing from the signature of a uniform temperature
spot only over very short timescales.

\subsection{Spot `imaging'}

Since the microlensing lightcurve is essentially a convolution integral
over the extended source, it is not surprising to find that the signatures 
of spots are non-unique, in the sense that spots of different shape and
temperature structure can give rise to light curves which are identical 
within observational error. Thus, the use of microlensing as a tool for
resolving detailed maps of the stellar photosphere is highly limited.
Nevertheless, since the effect of each spot is fairly localised, one can at 
least use the light curve to place constraints on spot position and `filling
factor'. Similarly, a \emph{failure} to detect spot signatures from high
time resolution observations of a transit event -- particularly from e.g.
a fold caustic crossing -- can provide useful limits on surface activity for
stellar atmosphere models.

\subsection{Estimating stellar radii}

The presence of starspots can also affect the estimation of the source
parameters; specifically, neglecting the effects of maculae in the 
lightcurve can produce a biased determination of the stellar radius, and 
thus the lens Einstein radius and proper motion. For example, a star with a 
central hot spot will give a best fit to an unspotted star of smaller stellar 
radius; moreover, the impact of sparse sampling may `mask' the presence of
the spot feature and thus yield a perfectly acceptable $\chi^2$ to the smaller
radius. The best fitting stellar radius estimated from V band data is 
typically 10\% smaller than the true radius, for a range of plausible spot 
parameters. However, at longer wavelengths (e.g. I and K bands) the bias in
the estimated radius is less severe, consistent with the reduced temperature
sensitivity of the Planck function at these wavelengths. Moreover, cool 
spots do not produce a strong bias in the estimated radius.

\subsection{Stellar rotation}

The area and temperature contrast of a spot essentially determine the shape
(i.e. amplitude and width) of the spot profile on the microlensing lightcurve.
Interestingly, for the case of a rotating source, one can change the ratio of
amplitude to width for the spot signature, since this effectively changes
the transverse velocity of the lens with respect to the \emph{spot}, without
changing the lens tranvserse velocity with respect to the star as a whole.
Whether the spot feature is narrowed or broadened depends on the orientation
of the lens trajectory with respect to the rotation axis of the source. Even
for the long rotation periods expected for late-type giants, the effect of
rotation can have an impact on the observed lightcurve -- e.g. for a source
rotation period $\tau_S = 0.05 t_e$ and a spot of radius of $15\deg$, the
rotation of the source may bias the estimation of the spot radius by up to
10\%. Of course a further complicating factor could be significant evolution
in the surface distribution of spots during the lensing event -- a problem
more likely to be relevant for faster rotators.

\begin{figure}
\plottwo{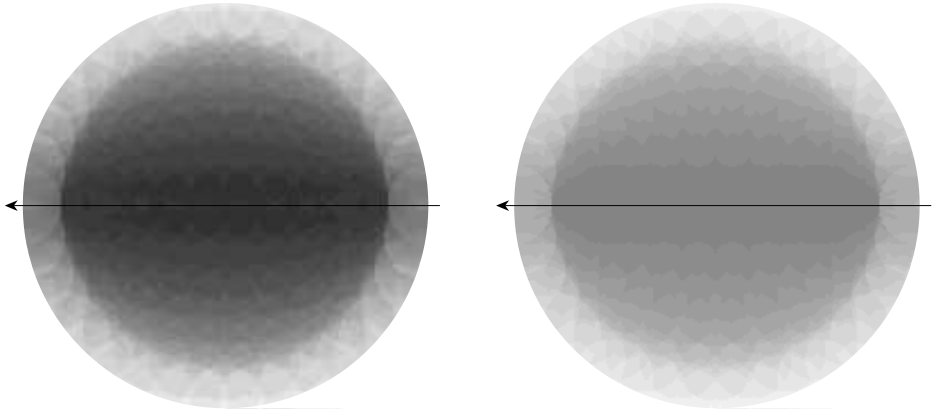}{helen5.ps}
\caption{\small{Grey scale map showing the detectability of a 10\deg
hot spot of temperature contrast $+1000$K, on a 0.1 AER star. A linear
limb darkening law is assumed. The arrow denotes the lens trajectory.
Shown are the results for V band (left panel) and R band (right panel)
observations.}}
\end{figure}

\begin{figure}
\plottwo{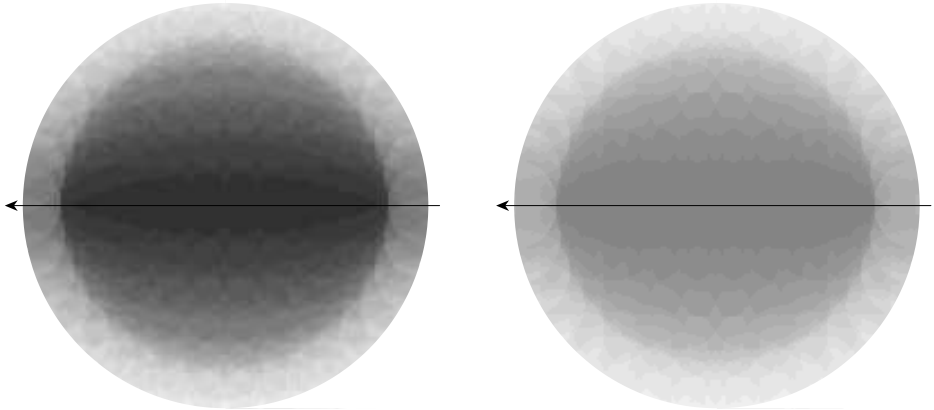}{helen7.ps}
\caption{\small{Same as for Fig. 3, but for a 0.01 AER star.}}
\end{figure}

\subsection{Maps of spot detectability}

As discussed above, the position of a spot on the stellar photosphere 
is the crucial factor which determines its detectability. As an 
illustration, Figures 3 and 4 show grey scale plots indicating the
detectability of a $10\deg$ spot, of temperature 5000K, at various 
positions on the photosphere of a star with $T_{eff} = 4000$K, 
$\log g = 2$. Regions where the spot provides a larger signature are 
darker; in this example the darkest level on the plots indicates an
average deviation of $\geq 10\%$ compared with the unspotted lightcurve,
while the lightest level indicates no significant deviation. Note that in 
the outer annulus of the stellar disk the spot is barely detectable,
due to the effects of limb darkening and geometrical foreshortening, although
this effect is more pronounced at high stellar latitudes, far from the lens
trajectory. Note also that spots are more detectable in the V band -- 
consistent with the greater bias in the determination of the stellar radius 
at shorter wavelengths.

\section{Small scale radial oscillations}
We have developed a model of the microlensing of stars displaying radial
oscillations. Although intrinsic variability is adopted as an
exclusion criterion by current microlensing surveys, this does not
of course preclude the possiblity that variable stars may themselves
be microlensed (c.f. EROS-2, Ansari \emph{et al.} 1995) -- 
and such an observation could provide 
important information on issues such as stellar pulsation and 
asteroseismology. Complex lightcurves are produced by the radial 
oscillations of even a simplified model -- see Fig. 5. Here we modelled 
a sinuosoidal oscillation only in the stellar radius, neglecting
chromatic variations other than those due to limb darkening. The period
of oscillation was 5 days, which would typically correspond to about 4 
complete cycles over the timescale of an event with small Einstein radius.
Higher order non-radial oscillations can be modelled as a series of spots;
however, detecting such features would be limited by the same criteria 
as discussed above -- with the additional possible limitation of evolution 
during the event. Note that, due to the effect of limb darkening, for a fixed
stellar radius the oscillation is more detectable at shorter wavelengths. 
Unlike spots, a small level of oscillation can be
detected without the need for a transit or near-transit event. For 
example, if the impact parameter $\leq 0.2 $ AER, a 2\% change in the
radius of a 0.001 AER radius star produces a highly significant deviation 
from the lightcurve modelled for a static star in the Bulge, for more than 
25\% of the event duration. For a well-sampled light curve, however, the 
deviations do not significantly bias the estimation of e.g. the stellar 
radius, although a poor or \emph{aliased} sampling rate could introduce 
such a bias. Thus, detection of small scale oscillations will only occur if 
aggressive temporal sampling strategies are employed.

\begin{figure}
\plotfiddle{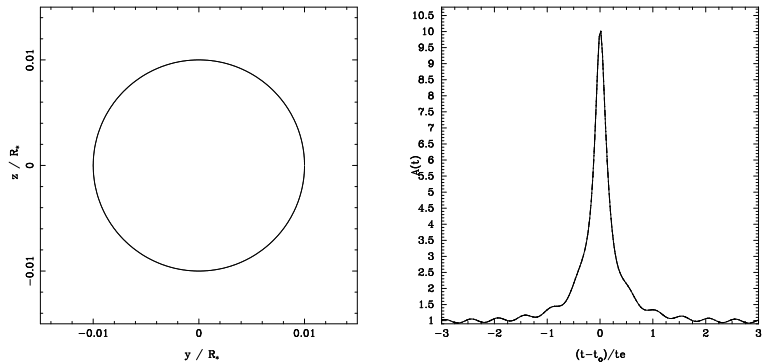}{5cm}{-90}{40}{40}{-150}{200}
\caption{\label{two}\small{The microlensing lightcurve produced by a
6000K, $\log g = 3$ star, varying in radius by 3\% from a mean radius of
0.01 AER being lensed with a minimum impact parameter 0.1. Most of the 
light curve gives a clear residual compared to the same star without the
oscillation. The colour bands are superimposed on each other in this
case as there is no strong chromatic effect for such a small stellar
radius.}}
\end{figure}

\section{Spectroscopic signatures}
Spectroscopic studies of transit microlensing events can provide
strong constraints on stellar atmosphere models, (c.f Heyrovsk\'{y}, 
Sasselov \& Loeb 2000). We expect to see equivalent widths of spectral
lines vary as the lens \emph{gravitationally images} the source. Strong 
resonant lines such as Ly$\alpha$, CaII and MgII appear 
brighter at the limb. Thus, in transit microlensing events spectroscopic 
targetting of such lines would yield good estimates of the stellar radius 
and hence the Einstein radius -- in a manner analogous to the polarimetric
signature modelled by Simmons \emph{et al.} (1995). Like the chromatic 
signature of extended source microlensing, the variation of line profile
shape produces a unique discriminant between microlensing and intrinsic
variability. To image spots, we need to probe lines sensitive to 
temperature structure: for example H$\alpha$ will be sensitive to active 
regions and molecular lines in the atmospheres of late-type giants will be
sensitive to the presence of hot spots. However the atmospheres of
red giants are poorly constrained by models: their limb darkening laws are 
extrapolated from limited data and are complicated by the presence of 
many molecules and by variations in metallicity. Moreover, many late-type
giants possess extended circumstellar envelopes, for which microlensing would
also present a powerful diagnostic tool. For example, the spectroscopic 
signature of a thin spherical shell of microlensed circumstellar material
can provide an important probe of the density and velocity structure of the
envelope (c.f. Ignace and Hendry 1999).

\section{Conclusions}

With the advent of automated `alert' status for candidate events, the 
prospects for using microlensing as a tool for gravitational imaging and
investigating stellar atmospheres are dramatically improved. Of crucial
importance in such studies are coordinated global observations to achieve
the high level of sampling required to constrain models effectively.
Observations of this quality are \emph{also} precisely what is required 
to extract
useful information from the microlensing of extended sources -- making
detailed modelling of such events a timely issue. Moreover, the 
intensive observation of second caustic crossings in binary events --
currently being pursued with the principal goal of detecting planetary
companions -- would also provide powerful constraints on the range of
spot-related phenomena discussed in this paper. We are currently, therefore, 
extending our analyses to consider the photometric, polarimetric and
spectroscopic signatures of star spots lensed during line caustic crossing
events.


\begin{references}

\reference Ansari, R. et al, 1995, \aap, {\bf{299}}, 21 
\reference Alcock, C. et al, 1997, \apj, {\bf{491}}, 436
\reference Claret and Gimenez, 1990, \apss, {\bf{169}}, 223
\reference Gould, A., 1994, \apj, {\bf{421}},71
\reference Hendry, M.A., et al , 1998, New Astron. Rev., {\bf{42}}, 125
\reference Hendry, M.A., Bryce, H.M. \& Valls-Gabaud, D., 2000, in prep.
\reference Heyrovsk\'{y}, D., Sasselov, D. and Loeb, A.2000, \apj,
submitted(astro-ph/9902273)
\reference Heyrovsk\'{y}, D. and Sasselov, D.2000, \apj, {\bf{529}}, 69
\reference Ignace, R. and Hendry, M. A., 1999, \aap, {\bf{341}}, 201
\reference Kim,H. et al, 2000, \mnras, submitted
\reference Simmons, J. F. L. et al, 1995, \mnras, {\bf{276}}, 182
\reference Uitenbroek, H., Dupree, A.K. \& Gilliland, R.L., 1998, \apj,
{\bf{116}}, 2501
\reference Valls-Gabaud, D., 1998., \mnras, {\bf{294}}, 747
\reference Witt, H.J. and Mao, S., 1994, \apj, {\bf{430}}, 50


\end{references}
\end{document}